\documentstyle[11pt,appb,epsf]{article} 
\begin{document}
\title{THE KINEMATICS AND DYNAMICS \\ OF FLAVOR MIXING\thanks{Invited
talk given at the XXI School of Theoretical Physics -- ``Recent Progress in
Theory and Phenomenology of Fundamental Interactions'', Ustro\a'{n},
Poland (1997).}}
% you can use \\ to break lines

\author{Harald Fritzsch
\address{Theory Division, CERN, CH--1211 Geneva 23, Switzerland \\
and \\
Sektion Physik, Universit\"at M\"unchen, D--80333 M\"unchen, Germany}
}
\maketitle
\begin{abstract}
In view of the observed strong hierarchy of the quark and lepton masses and
of the flavor mixing angles, it is argued that the description of flavor
mixing must take this into account. One particularly interesting way to describe
the flavor mixing, which, however, is not the one used today, emerges, which
is particularly suited for models of quark mass matrices based on flavor
symmetries. We conclude that the unitarity triangle important for $B$
physics should be close to or identical to a rectangular triangle. $CP$
violation is maximal in this sense.
\end{abstract}

The phenomenon of flavor mixing, which is intrinsically linked to
$CP$ violation, is an important ingredient of the Standard Model of Basic
Interactions. Yet unlike other features of the Standard Model, e.\ g.\ the
mixing of the neutral electroweak gauge bosons, it is a phenomenon which can
merely be described. A deeper understanding is still lacking, but most
theorists would agree that it is directly linked to the mass spectrum of
the quarks -- the possible mixing of lepton flavors will not be discussed
here. Furthermore there is a general consensus that a deeper dynamical
understanding would require to go beyond the physics of the Standard Model.
In this talk I shall not go thus far. Instead I shall demonstrate that the
observed properties of the flavor mixing, combined with our knowledge about
the quark mass spectrum, suggest specific symmetry properties which allow
to fix the flavor mixing parameters with high precision, thus predicting the
outcome of the experiments which will soon be performed at the $B$--meson
factories.

%\section{Text}
In the standard electroweak theory, the phenomenon of flavor
mixing of the quarks is described by a $3\times 3$ unitary matrix, the
Cabibbo-Kobayashi-Maskawa (CKM) matrix \cite{Cabibbo63,KM73}. This
matrix can be expressed in terms of four parameters, which are usually 
taken as three rotation angles and one phase. A number of
different parametrizations have been proposed in the literature
\cite{KM73}--\cite{FX97}. Of course, adopting a particular parametrization
of flavor mixing is arbitrary and not directly a physical
issue. Nevertheless it is quite likely that the actual values of 
flavor mixing parameters (including the strength of $CP$ violation),
once they are known with high precision, will give interesting information 
about the physics beyond the standard model. Probably at this point it 
will turn out that a particular description of the CKM matrix is more
useful and transparent than the others. For this reason, let me first 
analyze all possible parametrizations and point out their
respective advantages and disadvantages.

In the standard model the quark flavor mixing arises once the up- and
down-type mass matrices are diagonalized. The generation of quark masses 
is intimately related to the phenomenon of flavor mixing. In
particular, the flavor mixing parameters do depend on the
elements of quark mass matrices. A particular structure of the
underlying mass matrices calls for a particular choice of the
parametrization of the flavor mixing matrix. For example, in
Ref. \cite{Fritzsch79} it was noticed that a rather special form of
the flavor mixing matrix results, if one starts from Hermitian mass
matrices in which the (1,3) and (3,1) elements vanish. This has been
subsequently observed again in a number of papers
\cite{Hall}. Recently we have studied the exact form of such a
description from a general point of view and pointed out some
advantages of this type of representation in the discussion of flavor
mixing and $CP$-violating phenomena \cite{FX97}, which will be discussed
later.

In the standard model the weak charged currents are given by 
\begin{equation}
\overline{(u, ~ c, ~ t)}^{~}_L \left ( \matrix{
V_{ud}  & V_{us}        & V_{ub} \cr
V_{cd}  & V_{cs}        & V_{cb} \cr 
V_{td}  & V_{ts}        & V_{tb} \cr} \right ) 
\left ( \matrix{
d \cr s \cr b \cr} \right  )_L \; ,
%               (1)
\end{equation}
where $u$, $c$, ..., $b$ are the quark mass eigenstates, $L$ denotes
the left-handed fields, and $V_{ij}$ are elements of the CKM matrix
$V$. In general $V_{ij}$ are complex numbers, but their absolute
values are measurable quantities. For example, $|V_{cb}|$ primarily
determines the lifetime of $B$ mesons. The phases of $V_{ij}$,
however, are not physical, like the phases of quark fields. A phase
transformation of the $u$ quark ($u \rightarrow u ~ e^{{\rm
i}\alpha}$), for example, leaves the quark mass term invariant but
changes the elements in the first row of $V$ (i.e., $V_{uj} \rightarrow 
V_{uj} ~ e^{-{\rm i}\alpha}$). Only a common phase transformation of all 
quark fields leaves all elements of $V$ invariant, thus there is a
five-fold freedom to adjust the phases of $V_{ij}$.

In general the unitary matrix $V$ depends on nine parameters.
Note that in the absence of complex phases $V$ would consist of only three 
independent parameters, corresponding to three (Euler) rotation
angles. Hence one can describe the complex matrix $V$ by three
angles and six phases. Due to the freedom in redefining the quark
field phases, five of the six phases in $V$ can be absorbed and we arrive
at the well-known result that the CKM matrix $V$ can be parametrized
in terms of three rotation angles and one $CP$-violating phase. The
question about how many different ways to describe $V$ may exist was
raised some time ago \cite{Jarlskog89}. Below we shall
reconsider this problem and give a complete analysis.

If the flavor mixing matrix $V$ is first assumed to be a real orthogonal matrix, it can
in general be written as a product of three matrices $R_{12}$,
$R_{23}$ and $R_{31}$, which describe simple rotations in the (1,2),
(2,3) and (3,1) planes:
\begin{eqnarray}
R_{12}(\theta) & = & \left ( \matrix{
c^{~}_{\theta}  & s^{~}_{\theta}        & 0 \cr
- s^{~}_{\theta}        & c^{~}_{\theta}        & 0 \cr
0       & 0     & 1 \cr} \right ) \; , \nonumber \\ \nonumber \\
R_{23}(\sigma) & = & \left ( \matrix{
1       & 0     & 0 \cr
0       & c_{\sigma}    & s_{\sigma} \cr
0       & - s_{\sigma}  & c_{\sigma} \cr} \right ) \; , \nonumber \\
\nonumber \\
R_{31}(\tau) & = & \left ( \matrix{
c_{\tau}        & 0     & s_{\tau} \cr
0       & 1     & 0 \cr
- s_{\tau}      & 0     & c_{\tau} \cr} \right ) \; ,
%               (2)
\end{eqnarray}
where $s^{~}_{\theta} \equiv \sin \theta$, $c^{~}_{\theta} \equiv \cos
\theta$, etc.
Clearly these rotation matrices do not commute with each other.
There exist twelve different ways to arrange products of these
matrices such that the most general orthogonal matrix $R$ can be
obtained \cite{Jarlskog89}. 
Note that the matrix $R^{-1}_{ij} (\omega) $ plays an equivalent role
as $R_{ij} (\omega) $ in constructing $R$, because of $R^{-1}_{ij}(\omega) =
R_{ij}(-\omega)$. Note also that $R_{ij} (\omega) R_{ij}
(\omega^{\prime}) = R_{ij} (\omega + \omega^{\prime})$ holds, thus 
the product $R_{ij}(\omega) R_{ij}(\omega^{\prime})
R_{kl}(\omega^{\prime\prime})$ or $R_{kl}(\omega^{\prime\prime})
R_{ij}(\omega) R_{ij}(\omega^{\prime})$ cannot cover the whole space
of a $3\times 3$ orthogonal matrix and should be excluded.
Explicitly the twelve different forms of $R$ read as
\begin{eqnarray}
(1) & & R \; =\; R_{12}(\theta) ~ R_{23}(\sigma) ~ R_{12}(\theta^{\prime})
\; , \nonumber \\
(2) & & R \; =\; R_{12}(\theta) ~ R_{31}(\tau) ~ R_{12}(\theta^{\prime})
\; , \nonumber \\
(3) & & R \; =\; R_{23}(\sigma) ~ R_{12}(\theta) ~ R_{23}(\sigma^{\prime})
\; , \nonumber \\
(4) & & R \; =\; R_{23}(\sigma) ~ R_{31}(\tau) ~ R_{23}(\sigma^{\prime})
\; , \nonumber \\
(5) & & R \; =\; R_{31}(\tau) ~ R_{12}(\theta) ~ R_{31}(\tau^{\prime})
\; , \nonumber \\
(6) & & R \; =\; R_{31}(\tau) ~ R_{23}(\sigma) ~ R_{31}(\tau^{\prime})
\; , \nonumber
\end{eqnarray}
in which a rotation in the $(i,j)$ plane occurs twice;
and
\begin{eqnarray}
(7) & & R \; =\; R_{12}(\theta) ~ R_{23}(\sigma) ~ R_{31}(\tau)
\; , \nonumber \\
(8) & & R \; =\; R_{12}(\theta) ~ R_{31}(\tau) ~ R_{23}(\sigma)
\; , \nonumber \\
(9) & & R \; =\; R_{23}(\sigma) ~ R_{12}(\theta) ~ R_{31}(\tau)
\; , \nonumber \\
(10) & & R \; =\; R_{23}(\sigma) ~ R_{31}(\tau) ~ R_{12}(\theta)
\; , \nonumber \\
(11) & & R \; =\; R_{31}(\tau) ~ R_{12}(\theta) ~ R_{23}(\sigma)
\; , \nonumber \\
(12) & & R \; =\; R_{31}(\tau) ~ R_{23}(\sigma) ~ R_{12}(\theta) 
\; , \nonumber 
\end{eqnarray}
where all three $R_{ij}$ are present. 

Although all the above twelve combinations represent the most general 
orthogonal matrices, only nine of them are structurally different.
The rea-son is that the products $R_{ij} R_{kl} R_{ij}$ and $R_{ij} R_{mn} R_{ij}$ (with
$ij\neq kl\neq mn$) are correlated with each other, leading
essentially to the same form for $R$. Indeed it is straightforward to
see the correlation between patterns (1), (3), (5) and (2), (4),
(6), respectively, as follows:
\begin{eqnarray}
R_{12}(\theta) ~ R_{31}(\tau) ~ R_{12}(\theta^{\prime})
& = & R_{12}(\theta + \pi/2) ~ R_{23}(\sigma = \tau) ~
R_{12}(\theta^{\prime} - \pi/2) \; , \nonumber \\
R_{23}(\sigma) ~ R_{31}(\tau) ~ R_{23}(\sigma^{\prime})
& = & R_{23}(\sigma -\pi/2) ~ R_{12}(\theta = \tau) ~
R_{23}(\sigma^{\prime} + \pi/2) \; , \nonumber \\
R_{31}(\tau) ~ R_{23}(\sigma) ~ R_{31}(\tau^{\prime})
& = & R_{31}(\tau  + \pi/2) ~ R_{12}(\theta = \sigma) ~
R_{31}(\tau^{\prime} - \pi/2) \; .
%               (3)
\end{eqnarray}
Thus the orthogonal matrices (2), (4) and (6) need not be treated as
independent choices. 
We then draw the conclusion that
there exist {\it nine} different forms for the orthogonal matrix $R$,
i.e., patterns (1), (3) and (5) as well as (7) -- (12). 

We proceed to include the $CP$-violating phase, denoted by $\varphi$,
in the above rotation matrices. The resultant matrices should be
unitary such that a unitary flavor mixing matrix can be finally
produced. There are several different ways for
$\varphi$ to enter $R_{12}$, e.g., 
$$
R_{12} (\theta, \varphi) \; =\; \left ( \matrix{
c^{~}_{\theta}  & s^{~}_{\theta} ~ e^{+{\rm i} \varphi} & 0 \cr
- s^{~}_{\theta} ~ e^{-{\rm i} \varphi}         & c^{~}_{\theta}        & 0 \cr
0       & 0     & 1 \cr} \right ) \; , 
\eqno(4{\rm a})
%               (4a)
$$
or
$$
R_{12} (\theta, \varphi) \; =\; \left ( \matrix{
c^{~}_{\theta}  & s^{~}_{\theta}        & 0 \cr
- s^{~}_{\theta}        & c^{~}_{\theta}        & 0 \cr
0       & 0     & e^{-{\rm i} \varphi} \cr} \right ) \; , 
\eqno(4{\rm b})
%               (4b)
$$
or
$$
R_{12} (\theta, \varphi) \; =\; \left ( \matrix{
c^{~}_{\theta} ~ e^{+{\rm i} \varphi}   & s^{~}_{\theta}        & 0 \cr
- s^{~}_{\theta}        & c^{~}_{\theta} ~ e^{-{\rm i} \varphi} & 0 \cr
0       & 0     & 1 \cr} \right ) \; . 
\eqno(4{\rm c})
%               (4c)
$$
Similarly one may introduce a phase parameter into $R_{23}$ or
$R_{31}$. Then the CKM matrix $V$ can be constructed, as a product of
three rotation matrices, by use of one complex $R_{ij}$ and two real ones. 
Note that the location of the $CP$-violating phase in $V$ can be arranged by
redefining the quark field phases, thus it does not play an essential role in 
classifying different parametrizations. We find that it is always
possible to locate the phase parameter $\varphi$ in a $2\times 2$ submatrix of
$V$, in which each element is a sum of two terms with the relative
phase $\varphi$. The remaining five elements of $V$ are real in such a 
phase assignment. Accordingly we arrive at the nine distinctive
parametrizations of the CKM matrix $V$ listed in Table 1 where
the complex rotation matrices $R_{12}(\theta, \varphi)$,
$R_{23}(\sigma, \varphi)$ and $R_{31}(\tau, \varphi)$ are obtained
directly from the real ones in Eq. (2) with the replacement $1
\rightarrow e^{-{\rm i}\varphi}$. These nine possibilities have been
discussed recently in Ref. \cite{FritzschXing97} (see also Ref. \cite{Rasin97}).

One can see that {\it P2} and {\it P3}
correspond to the Kobayashi-Maskawa \cite{KM73} and Maiani
\cite{Others} representations, although different
notations for the $CP$-violating phase and three mixing angles are adopted
here. The latter is indeed equivalent to the 
``standard'' parametrization advocated by the Particle Data Group
\cite{Others,GKR}. This can be seen clearly if one makes 
three transformations of quark field phases: 
$c \rightarrow c ~ e^{-{\rm i} \varphi}$, $t
\rightarrow t ~ e^{-{\rm i} \varphi}$, and $b \rightarrow 
b ~ e^{-{\rm i} \varphi}$. In addition, {\it P1} is just the one discussed
by Xing and me in Ref. \cite{FX97}.

{\rm From} a mathematical point of view, all nine different parametrizations
are equivalent. However this is not the case if we apply our
considerations to the quarks and their mass spectrum. It is well--known that
both the observed
quark mass spectrum and the observed values of the flavor mixing
parameters exhibit a striking hierarchical structure. The latter can
be understood in a natural way as the consequence of a specific
pattern of chiral symmetries whose breaking causes the towers of
different masses to appear step by step
\cite{Fritzsch87a,Fritzsch87b,Hall93a}. Such a chiral evolution of the
mass matrices leads, as argued in Ref. \cite{Fritzsch87b}, to a
specific way to introduce and describe the flavor mixing. In the limit 
$m_u = m_d =0$, which is close to the real world, since $m_u/m_t \ll 1$ and
$m_d/m_b \ll 1$, the flavor mixing is merely a rotation between the
$t$--$c$ and $b$--$s$ systems, described by one rotation angle. No complex
phase is present; i.e., $CP$ violation is absent. This rotation angle
is expected to change very little, once $m_u$ and $m_d$
are introduced as tiny perturbations. A sensible parametrization should
make use of this feature. This implies that the rotation matrix
$R_{23}$ appears exactly once in the description of the CKM matrix
$V$, eliminating {\it P2} (in which $R_{23}$ appears twice) and {\it P5}
(where $R_{23}$ is absent). This leaves us with seven parametrizations 
of the flavor mixing matrix.

The list can be reduced further by considering the location of the phase $\varphi$. In
the limit $m_u = m_d =0$, the phase must disappear in the weak
transition elements $V_{tb}$, $V_{ts}$, $V_{cb}$ and $V_{cs}$. In
{\it P7} and {\it P8}, however, $\varphi$ appears particularly in
$V_{tb}$. Thus these two parametrizations should be eliminated, leaving
us with five parametrizations (i.e., {\it P1}, {\it P3}, {\it P4}, {\it P6} and
{\it P9}). In the same limit, the phase $\varphi$ appears in the $V_{ts}$
element of {\it P3} and the $V_{cb}$ element of {\it P4}. Hence these two
parametrizations should also be eliminated. Then we are left with three
parametrizations, {\it P1}, {\it P6} and {\it P9}. As expected, these are the
parametrizations containing the complex rotation matrix
$R_{23}(\sigma, \varphi)$. We stress that the ``standard'' parametrization
\cite{GKR} (equivalent to {\it P3}) does not obey the above constraints and
should be dismissed.

Among the remaining three parametrizations, {\it P1} is singled out by 
the fact that the $CP$-violating phase $\varphi$ appears only in the
$2\times 2$ submatrix of $V$ describing the weak transitions among the 
light quarks. This is precisely the phase where the phase $\varphi$
should appear, not in any of the weak transition elements involving the 
heavy quarks $t$ and $b$.

In the parametrization {\it P6} or {\it P9}, the complex
phase $\varphi$ appears in $V_{cb}$ or $V_{ts}$, but this phase factor
is multiplied by a product of $\sin\theta$ and $\sin\tau$, i.e., it is of second 
order of the weak mixing angles. Hence the imaginary parts of these
elements are not exactly vanishing, but very small in magnitude. 

In our view the best possibility to describe the flavor mixing in the
standard model is to adopt the parametrization {\it P1}. As discussed
in Ref. \cite{FX97}, this parametrization has a number of significant
advantages in addition to that mentioned above. Especially it is well
suited for specific models of quark mass matrices. 

In the following part I shall show that the parametrization {\it P1} follows
automatically if we impose the constraints from the chiral symmetries and the
hierarchical structure of the mass eigenvalues. We take the point of view
that the quark mass eigenvalues are dynamical entities, and one could change
their values in order to study certain
symmetry limits, as it is done in QCD. In the standard electroweak model, in
which the quark mass matrices are given by the coupling of a scalar field to
various quark fields, this can certainly be done by adjusting the related
coupling constants. Whether it is possible in reality is an open question. It
is well--known that the quark mass matrices can always be made hermitian
by a suitable transformation of the right--handed fields. Without loss of
generality, we shall suppose in this paper that the quark mass matrices are
hermitian. In the limit where the masses of the $u$ and $d$ quarks are set to
zero, the quark mass matrix $\tilde{M}$ (for both charge $+2/3$ and
charge $-1/3$ sectors) can be arranged such that its elements 
$\tilde{M}_{i1}$ and $\tilde{M}_{1i}$ ($i=1,2,3$) are all zero 
\cite{Fritzsch87a,Fritzsch87b}. Thus the quark
mass matrices have the form
\setcounter{equation}{4}
\begin{equation}
\tilde{M} \; =\; \left ( \matrix{
0       & 0     & 0 \cr
0       & \tilde{C}     & \tilde{B} \cr
0       & \tilde{B}^*   & \tilde{A} \cr} \right ) \; .
%               (5)
\end{equation}
The observed mass hierarchy is incorporated into this structure by
denoting the entry which is of the order of the $t$-quark or 
$b$-quark mass by $\tilde{A}$, with $\tilde{A}\gg \tilde{C},
|\tilde{B}|$. It can easily be seen (see, e.g., Ref. \cite{Lehmann96}) that
the complex phases in the mass matrices (5) can be
rotated away by subjecting both $\tilde{M}_{\rm u}$ and $\tilde{M}_{\rm d}$ to the
same unitary transformation. Thus we shall take $\tilde{B}$ to be
real for both up- and down-quark sectors. As expected, $CP$ violation
cannot arise at this stage. The diagonalization of the mass matrices
leads to a mixing between the second and third families, described by an
angle $\tilde{\theta}$. The flavor mixing matrix is
then given by
\begin{equation}
\tilde{V} \; =\; \left ( \matrix{
1       & 0     & 0 \cr
0       & \tilde{c}     & \tilde{s} \cr
0       & -\tilde{s}    & \tilde{c} \cr } \right ) \; ,
%               (6)
\end{equation}
where $\tilde{s} \equiv \sin \tilde{\theta}$ and $\tilde{c} \equiv
\cos \tilde{\theta}$. In view of the fact that the limit $m_u = m_d
=0$ is not far from reality, the angle $\tilde{\theta}$ is essentially 
given by the observed value of $|V_{cb}|$ ($=0.039 \pm 0.002$
\cite{Neubert96,Forty97});
i.e., $\tilde{\theta} = 2.24^{\circ} \pm 0.12^{\circ}$. 

At the next and final stage of the chiral evolution of the mass matrices,
the masses of the $u$ and $d$ quarks are introduced.
The Hermitian mass matrices have in general the
form:
\begin{equation}
M \; =\; \left ( \matrix{
E       & D     & F \cr
D^*     & C     & B \cr
F^*     & B^*   & A \cr } \right ) \; 
%               (7)
\end{equation}
with $A\gg C, |B| \gg E, |D|, |F|$. By a common unitary transformation of 
the up- and down-type quark fields, one can always arrange the mass
matrices $M_{\rm u}$ and $M_{\rm d}$ in such a way that $F_{\rm u} =
F_{\rm d} =0$; i.e.,
\begin{equation}
M \; =\; \left ( \matrix{
E       & D     & 0 \cr
D^*     & C     & B \cr
0       & B^*   & A \cr } \right ) \; .
%               (8)
\end{equation}
This can easily be seen as follows. If phases are neglected, the two
symmetric mass matrices $M_{\rm u}$ and $M_{\rm d}$ can be transformed 
by an orthogonal transformation matrix $O$, which can be described by
three angles such that they assume the form (8). The condition
$F_{\rm u} =F_{\rm d} =0$ gives two constraints for the three angles of 
$O$. If complex phases are allowed in $M_{\rm u}$ and $M_{\rm d}$, the 
condition $F_{\rm u} =F_{\rm u}^* = F_{\rm d} =F_{\rm d}^* =0$ imposes
four constraints, which can also be fulfilled, if $M_{\rm u}$ and
$M_{\rm d}$ are subjected to a common unitary transformation matrix $U$. The
latter depends on nine parameters. Three of them are not suitable for
our purpose, since they are just diagonal phases; but the remaining
six can be chosen such that the vanishing of $F_{\rm u}$ and $F_{\rm
d}$ results. 

The basis in which the mass matrices take the form (8) is a basis in
the space of quark flavors, which in our view is of special
interest. It is a basis in which the mass matrices exhibit two
texture zeros, for both up- and down-type quark sectors. 
These, however, do not imply special relations among
mass eigenvalues and flavor mixing parameters (as pointed out
above). In this basis the mixing is of the ``nearest neighbour'' form, 
since the (1,3) and (3,1) elements of $M_{\rm u}$ and $M_{\rm d}$
vanish; no direct mixing between the heavy $t$ (or $b$) quark and the
light $u$ (or $d$) quark is present.
In certain models (see, e.g., Refs. \cite{Fritzsch79,Hall}),
this basis is indeed of particular interest, but we shall proceed without 
relying on a special texture models for the mass matrices.

A mass matrix of the type (8) can in the absence of complex phases be
diagonalized by a rotation matrix, described only by two angles in the
hierarchy limit of quark masses \cite{Fritzsch79}.
At first the off-diagonal element 
$B$ is rotated away by a rotation between the second and third 
families (angle $\theta_{23}$); at the second step the element $D$ is rotated away by a
transformation of the first and second families (angle $\theta_{12}$). No rotation between
the first and third families is required 
\footnote{This is true in lowest order as the corresponding
rotation angle is small enough, in accordance with the hierarchical
structure discussed above. However, our final result of parametrizing
the flavor mixing matrix will remain exact, independent of the
mentioned approximation. The reason is that a rotation between the
first and third families can always be absorbed by redefining the
relevant rotation matrices. For a detailed discussion, see Ref. \cite{FritzschXing98}.}.
The rotation matrix for this sequence takes the form
\begin{eqnarray}
R \; =\; R_{12} R_{23} & = & \left ( \matrix{
c_{12}  & s_{12}        & 0 \cr
-s_{12} & c_{12}        & 0 \cr
0       & 0     & 1 \cr } \right )  \left ( \matrix{
1       & 0     & 0 \cr
0       & c_{23}        & s_{23} \cr
0       & -s_{23}       & c_{23} \cr } \right ) \; 
\nonumber \\ \nonumber \\
& = & \left ( \matrix{
c_{12}  & s_{12} c_{23} & s_{12} s_{23} \cr 
-s_{12} & c_{12} c_{23} & c_{12} s_{23} \cr
0       & -s_{23}       & c_{23} \cr } \right ) \; ,
%               (9)
\end{eqnarray}
where $c_{12} \equiv \cos \theta_{12}$, $s_{12} \equiv \sin
\theta_{12}$, etc.
The flavor mixing matrix $V$ is the product of two such matrices, one
describing the rotation among the up-type quarks, and the other describing
the rotation among the down-type quarks:
\begin{equation}
V \; =\; R^{\rm u}_{12} R^{\rm u}_{23} ( R^{\rm d}_{23} )^{-1} ( R^{\rm d}_{12} )^{-1} \; .
%               (10)
\end{equation}
The product $R^{\rm u}_{23} (R^{\rm d}_{23} )^{-1}$ can be written as
a rotation matrix described by a single angle $\theta$. In the limit
$m_u = m_d =0$, this is just the angle $\tilde{\theta}$ encountered
in Eq. (6). The angle which describes the $R^{\rm u}_{12}$ rotation shall
be denoted by $\theta_{\rm u}$; the corresponding angle for the
$R^{\rm d}_{12}$ rotation by $\theta_{\rm d}$. Thus in the absence of
$CP$-violating phases the flavor mixing matrix takes the following 
specific form:
\begin{eqnarray}
V & = & \left ( \matrix{
c_{\rm u}       & s_{\rm u}     & 0 \cr
-s_{\rm u}      & c_{\rm u}     & 0 \cr
0       & 0     & 1 \cr } \right )  \left ( \matrix{
1       & 0     & 0 \cr
0       & c     & s \cr
0       & -s    & c \cr } \right )  \left ( \matrix{
c_{\rm d}       & -s_{\rm d}    & 0 \cr
s_{\rm d}       & c_{\rm d}     & 0 \cr
0       & 0     & 1 \cr } \right )  \nonumber \\ \nonumber \\
& = & \left ( \matrix{
s_{\rm u} s_{\rm d} c + c_{\rm u} c_{\rm d}     & 
s_{\rm u} c_{\rm d} c - c_{\rm u} s_{\rm d}     & s_{\rm u} s \cr
c_{\rm u} s_{\rm d} c - s_{\rm u} c_{\rm d}     & 
c_{\rm u} c_{\rm d} c + s_{\rm u} s_{\rm d}     & c_{\rm u} s \cr
-s_{\rm d} s    & -c_{\rm d} s  & c \cr } \right ) \; ,
%               (11)
\end{eqnarray}
where $c_{\rm u} \equiv \cos\theta_{\rm u}$, $s_{\rm u} \equiv
\sin\theta_{\rm u}$, etc.

We proceed by including the phase parameters of the quark mass
matrices in Eq. (8). Each mass matrix has in general two complex 
phases. These phases can be dealt with in a similar way as described
in Refs. \cite{Fritzsch79,Fritzsch77}. It can easily be seen that, 
by suitable rephasing of the quark fields,
the flavor mixing matrix can finally be written in terms of only a
single phase $\varphi$ as follows \cite{FX97}:
\begin{eqnarray}
V & = & \left ( \matrix{
c_{\rm u}       & s_{\rm u}     & 0 \cr
-s_{\rm u}      & c_{\rm u}     & 0 \cr
0       & 0     & 1 \cr } \right )  \left ( \matrix{
e^{-{\rm i}\varphi}     & 0     & 0 \cr
0       & c     & s \cr
0       & -s    & c \cr } \right )  \left ( \matrix{
c_{\rm d}       & -s_{\rm d}    & 0 \cr
s_{\rm d}       & c_{\rm d}     & 0 \cr
0       & 0     & 1 \cr } \right )  \nonumber \\ \nonumber \\
& = & \left ( \matrix{
s_{\rm u} s_{\rm d} c + c_{\rm u} c_{\rm d} e^{-{\rm i}\varphi} &
s_{\rm u} c_{\rm d} c - c_{\rm u} s_{\rm d} e^{-{\rm i}\varphi} &
s_{\rm u} s \cr
c_{\rm u} s_{\rm d} c - s_{\rm u} c_{\rm d} e^{-{\rm i}\varphi} &
c_{\rm u} c_{\rm d} c + s_{\rm u} s_{\rm d} e^{-{\rm i}\varphi}   &
c_{\rm u} s \cr
- s_{\rm d} s   & - c_{\rm d} s & c \cr } \right ) \; .
%               (12)
\end{eqnarray}
Note that the three angles $\theta_{\rm u}$, $\theta_{\rm d}$ and
$\theta$ in Eq. (12) can all be arranged to lie in the first quadrant
through a suitable redefinition of quark field phases. Consequently
all $s_{\rm u}$, $s_{\rm d}$, $s$ and $c_{\rm u}$, $c_{\rm d}$, $c$
are positive. The phase $\varphi$ can in general take values from 0
to $2\pi$; and $CP$ violation is present in weak interactions
if $\varphi \neq 0, \pi$ and $2\pi$.

This particular and exact representation of the flavor mixing matrix was first
obtained by Xing and me in Ref. \cite{FX97}. In comparison with all other parametrizations discussed
previously \cite{KM73,Others,GKR}, it
has a number of interesting features which in our view make it very
attractive and provide strong arguments for its use in future
discussions of flavor mixing phenomena, in particular, those in
$B$-meson physics. We shall discuss them below.

a) The flavor mixing matrix $V$ in Eq. (12) follows directly from the
chiral expansion of the mass
matrices. Thus it naturally takes into account the hierarchical structure of the 
quark mass spectrum.

b) The complex phase describing $CP$ violation ($\varphi$) appears only in the
(1,1), (1,2), (2,1) and (2,2) elements of $V$, i.e., 
in the elements involving only the quarks of the first and second
families. This is a natural description of $CP$ violation since in our 
hierarchical approach $CP$ violation is not directly linked to the third family, but
rather to the first and second ones, and in particular to the mass terms of the
$u$ and $d$ quarks. 

It is instructive to consider the special case $s_{\rm u} = s_{\rm d}
= s = 0$. Then the flavor mixing matrix $V$ takes the form
\begin{equation}
V \; = \; \left ( \matrix{
e^{-{\rm i}\varphi}     & 0     & 0 \cr
0       & 1     & 0 \cr
0       & 0     & 1 \cr} \right ) \; .
%               (13)
\end{equation}
This matrix describes a phase change in the weak transition between
$u$ and $d$, while no phase change is present in the
transitions between $c$ and $s$ as well as $t$ and $b$.
Of course, this effect can be absorbed in a phase change of the $u$-
and $d$-quark fields, and no $CP$ violation is present. Once the
angles $\theta_{\rm u}$, $\theta_{\rm d}$ and $\theta$ are introduced, 
however, $CP$ violation arises. It is due to a phase change in the weak
transition between $u^{\prime}$ and $d^{\prime}$, where $u^{\prime}$
and $d^{\prime}$ are the rotated quark fields, obtained by applying
the corresponding rotation matrices given in Eq. (12) to the 
quark mass eigenstates ($u^{\prime}$: mainly $u$, small admixture of
$c$; $d^{\prime}$: mainly $d$, small admixture of $s$).

Since the mixing matrix elements involving $t$ or $b$ quark are real
in the representation (12), one can find that the phase parameter of
$B^0_q$-$\bar{B}^0_q$ mixing ($q=d$ or $s$), dominated by the
box-diagram contributions in the standard model \cite{PDG96}, is essentially unity:
\begin{equation}
\left ( \frac{q}{p} \right )_{B_q} = \;
\frac{V^*_{tb}V_{tq}}{V_{tb}V^*_{tq}} \; = \; 1 \; .
%               (14)
\end{equation}
In most of other parametrizations of the flavor mixing matrix,
however, the two rephasing-variant quantities 
$(q/p)^{~}_{B_d}$ and $(q/p)^{~}_{B_s}$ take different (maybe complex) values.

c) The dynamics of flavor mixing can easily be interpreted by
considering certain limiting cases in Eq. (12). In the limit $\theta
\rightarrow 0$ (i.e., $s \rightarrow 0$ and $c\rightarrow 1$), the
flavor mixing is, of course, just a mixing between the first and
second families, described by only one mixing angle (the Cabibbo angle 
$\theta_{\rm C}$ \cite{Cabibbo63}).  
It is a special and essential feature of the representation (12) that the Cabibbo
angle is {\it not} a basic angle, used in the parametrization. 
The matrix element $V_{us}$ (or $V_{cd}$) is
indeed a superposition of two terms including a phase. This feature
arises naturally in our hierarchical approach, but it is not new. In
many models of specific textures of mass matrices, it is indeed the
case that the Cabibbo-type transition $V_{us}$ (or $V_{cd}$) 
is a superposition of several
terms. At first, it was obtained by me 
in discussing the two-family mixing \cite{Fritzsch77}.

In the limit $\theta =0$ considered here, one has $|V_{us}| = |V_{cd}|
= \sin\theta_{\rm C} \equiv s^{~}_{\rm C}$ and
\begin{equation}
s^{~}_{\rm C} \; =\; \left | s_{\rm u} c_{\rm d} ~ - ~ c_{\rm u} s_{\rm d}
e^{-{\rm i}\varphi} \right | \; .
%               (15)
\end{equation}
This relation describes a triangle in the complex plane, as
illustrated in Fig. 1, which we shall denote as the ``light quark (LQ) 
triangle''. 
%%%%%%%%%%%%%%%%%%%%%%%%%% Fig 1 %%%%%%%%%%%%%%%%%%%
\begin{figure}[t]
\begin{picture}(400,160)(-30,210)
\put(80,300){\line(1,0){150}}
\put(80,300.5){\line(1,0){150}}
%\put(80,299.5){\line(1,0){150}}
\put(150,285.5){\makebox(0,0){$s^{~}_{\rm C}$}}
\put(80,300){\line(1,3){21.5}}
\put(80,300.5){\line(1,3){21.5}}
\put(80,299.5){\line(1,3){21.5}}
\put(70,335){\makebox(0,0){$s_{\rm u} c_{\rm d}$}}
\put(230,300){\line(-2,1){128}}
\put(230,300.5){\line(-2,1){128}}
%\put(230,299.5){\line(-2,1){128}}
\put(178,343.5){\makebox(0,0){$c_{\rm u} s_{\rm d}$}}

%\put(95,310){\makebox(0,0){$\gamma$}}
%\put(188,309){\makebox(0,0){$\beta$}}
\put(108,350){\makebox(0,0){$\varphi$}}
\end{picture}
\vspace{-2.7cm}
\caption{The light quark (LQ) triangle in the complex plane.}
\end{figure}
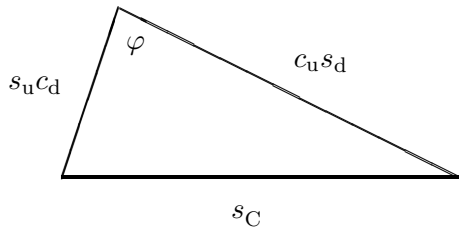
%%%%%%%%%%%%%%%%%%%%%%%%%%%%%%%%%%%%%%%%%%%%%%%%%
This triangle is a feature of
the mixing of the first two families (see also
Ref. \cite{Fritzsch77}). Explicitly one has (for $s=0$):
\begin{equation}
\tan\theta_{\rm C} \; =\; \sqrt{\frac{\tan^2\theta_{\rm u} +
\tan^2\theta_{\rm d} - 2 \tan\theta_{\rm u} \tan\theta_{\rm d}
\cos\varphi}
{1 + \tan^2\theta_{\rm u} \tan^2\theta_{\rm d} + 2 \tan\theta_{\rm u}
\tan\theta_{\rm d} \cos\varphi}} \; .
%               (16)
\end{equation}
Certainly the flavor mixing matrix $V$ cannot accommodate $CP$ violation in this
limit. However, the existence of $\varphi$ seems necessary in order
to make Eq. (16) compatible with current data, as one can see below.

d) The three mixing angles $\theta$, $\theta_{\rm u}$ and 
$\theta_{\rm d}$ have a precise physical meaning. The angle $\theta$
describes the mixing between the second and third families, which is
generated by the off-diagonal terms $B_{\rm u}$ and $B_{\rm d}$ in the 
up and down mass matrices of Eq. (8). 
We shall refer to this mixing involving $t$ and $b$ as the ``heavy
quark mixing''. The angle $\theta_{\rm u}$,
however, solely describes the $u$-$c$ mixing in the limit $m_u \ll m_c 
\ll m_t$, corresponding to the $D_{\rm
u}$ term in $M_{\rm u}$. We shall denote this as the ``u-channel mixing''.
The angle $\theta_{\rm d}$ solely describes 
the $d$-$s$ mixing in the limit $m_d \ll m_s \ll m_b$, 
corresponding to the $D_{\rm d}$ term in $M_{\rm
d}$; this will be denoted as the ``d-channel mixing''. 
Thus there exists an asymmetry between the mixing of the first and
second families and that of the second and third families,
which in our view reflects interesting details of the underlying dynamics of
flavor mixing. 
The heavy quark mixing is a combined effect, involving both charge
$+2/3$ and charge $-1/3$ quarks, while the u- or d-channel mixing
(described by the angle $\theta_{\rm u}$ or $\theta_{\rm d}$) proceeds 
solely in the charge $+2/3$ or charge $-1/3$ sector. Therefore an
experimental determination of these two angles would allow to draw
interesting conclusions about the amount and perhaps the underlying
pattern of the u- or d-channel mixing.

e) The three angles $\theta$, $\theta_{\rm u}$ and $\theta_{\rm d}$
are related in a very simple way to observable quantities of $B$-meson 
physics. 
For example, $\theta$ is related to 
the rate of the semileptonic decay $B\rightarrow D^*l\nu^{~}_l$; 
$\theta_{\rm u}$ is associated with the ratio of the decay rate of
$B\rightarrow (\pi, \rho) l \nu^{~}_l$ to that of $B\rightarrow 
D^* l\nu^{~}_l$; and $\theta_{\rm d}$ can be determined from the ratio of
the mass difference between two $B_d$ mass eigenstates to that between
two $B_s$ mass eigenstates. From Eq. (12) we find the following exact
relations:
\begin{equation}
\sin \theta \; = \; |V_{cb}| \sqrt{ 1 + \left |\frac{V_{ub}}{V_{cb}}
\right |^2} \; ,
%               (17)
\end{equation}
and
\begin{eqnarray}
\tan\theta_{\rm u} & = & \left | \frac{V_{ub}}{V_{cb}} \right | \; ,
\nonumber \\
\tan\theta_{\rm d} & = & \left | \frac{V_{td}}{V_{ts}} \right | \; .
%               (18)
\end{eqnarray}
These simple results make the parametrization (12) uniquely favorable 
for the study of $B$-meson physics.

By use of current data on $|V_{ub}|$ and $|V_{cb}|$, i.e., $|V_{cb}| = 
0.039 \pm 0.002$ \cite{Neubert96,Forty97} and $|V_{ub}/V_{cb}| =0.08 \pm 0.02$ 
\cite{PDG96}, we obtain $\theta_{\rm u} = 4.57^{\circ} \pm
1.14^{\circ}$ and $\theta = 2.25^{\circ} \pm 0.12^{\circ}$. Taking
$|V_{td}| = (8.6 \pm 2.1) \times 10^{-3}$ \cite{Forty97},
which was obtained from the analysis of current data on
$B^0_d$-$\bar{B}^0_d$ mixing,
we get $|V_{td}/V_{ts}| = 0.22 \pm 0.07$, i.e., $\theta_{\rm d} = 12.7^{\circ} 
\pm 3.8^{\circ}$.
Both the heavy quark mixing angle $\theta$ and the u-channel mixing
angle $\theta_{\rm u}$ are relatively small. The smallness of $\theta$ 
implies that Eqs. (15) and (16) are valid to a high degree of
precision (of order $1-c \approx 0.001$).

f) According to Eq. (16), as well as Eq. (15), the phase $\varphi$ is
a phase difference between the contributions to $V_{us}$ (or $V_{cd}$) 
from the u-channel mixing and the d-channel mixing. Therefore
$\varphi$ is given by the relative phase of $D_{\rm d}$ and $D_{\rm
u}$ in the quark mass matrices (8), if the phases of $B_{\rm u}$ and
$B_{\rm d}$ are absent or negligible. 

The phase $\varphi$ is not likely to be $0^{\circ}$ or $180^{\circ}$, according
to the experimental values given above, even though the measurement of 
$CP$ violation in $K^0$-$\bar{K}^0$ mixing \cite{PDG96} is not taken
into account. For $\varphi =0^{\circ}$, one
finds $\tan\theta_{\rm C} = 0.14 \pm 0.08$; and for $\varphi =
180^{\circ}$, one gets $\tan\theta_{\rm C} = 0.30 \pm 0.08$. Both
cases are barely consistent with the value of $\tan\theta_{\rm
C}$ obtained from experiments ($\tan\theta_{\rm C} \approx
|V_{us}/V_{ud}| \approx 0.226$). 

g) The $CP$-violating phase $\varphi$ in the flavor mixing matrix $V$ can be
determined from $|V_{us}|$ ($= 0.2205 \pm 0.0018$ \cite{PDG96})
through the following formula, obtained easily from Eq. (12):
\begin{equation}
\varphi \; =\; \arccos \left ( \frac{s^2_{\rm u} c^2_{\rm d} c^2 +
c^2_{\rm u} s^2_{\rm d} - |V_{us}|^2}{2 s_{\rm u} c_{\rm u} s_{\rm d}
c_{\rm d} c} \right ) \; .
%               (19)
\end{equation}
The two-fold ambiguity associated with the value of $\varphi$, coming
from $\cos\varphi = \cos (2\pi - \varphi)$, is removed if one
takes $\sin\varphi >0$ into account (this is required by current data on
$CP$ violation in $K^0$-$\bar{K}^0$ mixing (i.e., $\epsilon^{~}_K$)
\cite{PDG96}). More precise measurements of the angles $\theta_{\rm u}$ and
$\theta_{\rm d}$ in the forthcoming experiments of $B$ physics will
remarkably reduce the uncertainty of $\varphi$ to be determined from Eq.
(19). This approach is of course complementary to the direct determination of
$\varphi$ from $CP$ asymmetries in some weak $B$-meson decays into hadronic
$CP$ eigenstates.

Taking the presently known phenomenological constraints on quark
mixing and $CP$ violation into account, we find that 
the value of $\varphi$ is most likely in the range $40^{\circ}$ to
$120^{\circ}$; the central value is $\varphi \approx 81^{\circ}$ \cite{FX97}. 
Note that $\varphi$ is essentially independent of the angle $\theta$,
due to the tiny observed value of the latter. 
Once $\tan\theta_{\rm d}$ is precisely measured, one shall be able to fix the
magnitude of $\varphi$ to a satisfactory degree of accuracy.

h) It is well known that $CP$ violation in the flavor mixing matrix $V$ can
be rephasing-invariantly described by a universal quantity ${\cal J}$
\cite{Jarlskog85}:
\begin{equation}
{\rm Im} \left( V_{il} V_{jm} V^*_{im} V^*_{jl} \right) \; = \; {\cal J}
\sum\limits^{3}_{k,n=1} \left( \epsilon_{ijk}\epsilon_{lmn} \right )
\, .
%               (20)
\end{equation}
In the parametrization (12), ${\cal J}$ reads
\begin{equation}
{\cal J} \; = \; s_{\rm u} c_{\rm u} s_{\rm d} c_{\rm d} s^2 c \sin
\varphi \; .
%               (21)
\end{equation}
Obviously $\varphi = 90^{\circ }$ leads to the maximal value of ${\cal J}$.
Indeed $\varphi =90^{\circ}$, a particularly interesting case for $CP$ 
violation, is quite consistent with
current data. One can see from Fig. 2 of Ref. \cite{FX97} that this possibility exists
if $0.202 \leq \tan\theta_{\rm d} 
\leq 0.222$, or $11.4^{\circ} \leq \theta_{\rm d} \leq 12.5^{\circ}$.
In this case the mixing term
$D_{\rm d}$ in Eq. (8) can be taken to be real, and the term $D_{\rm 
u}$ to be imaginary, if ${\rm Im}(B_{\rm u}) = {\rm Im} (B_{\rm d})
=0$ is assumed (see also Refs. \cite{Lehmann96,FX95}). 
Since in our description of the flavor mixing the
complex phase $\varphi$ is related in a simple way to the phases of
the quark mass terms, the case $\varphi = 90^{\circ}$ is especially
interesting. It can hardly be an accident, and this case should be
studied further. The possibility that the phase $\varphi$ describing
$CP$ violation in the standard model is given by the algebraic number
$\pi/2$ should be taken seriously. It may provide a useful clue
towards a deeper understanding of the origin of $CP$ violation
and of the dynamical origin of the fermion masses.

In Ref. \cite{FX95} the case $\varphi =90^{\circ}$ has been
denoted as ``maximal'' $CP$ violation. It implies in our framework 
that in the complex
plane the u-channel and d-channel mixings are perpendicular to each
other. In this special case (as well as $\theta\rightarrow 0$), we have 
\begin{equation}
\tan^2\theta_{\rm C} \; =\; \frac{\tan^2\theta_{\rm u} ~ + ~
\tan^2\theta_{\rm d}}{1 ~ + ~ \tan^2\theta_{\rm u} \tan^2\theta_{\rm
d}} \; .
%               (22)
\end{equation}
To a good approximation (with the relative error $\sim 2\%$), 
one finds $s^2_{\rm C} \approx s^2_{\rm u} + s^2_{\rm d}$. 
%%%%%%%%%%%%%%%%%%%% Fig. 2 %%%%%%%%%%%%%%%%%%%%
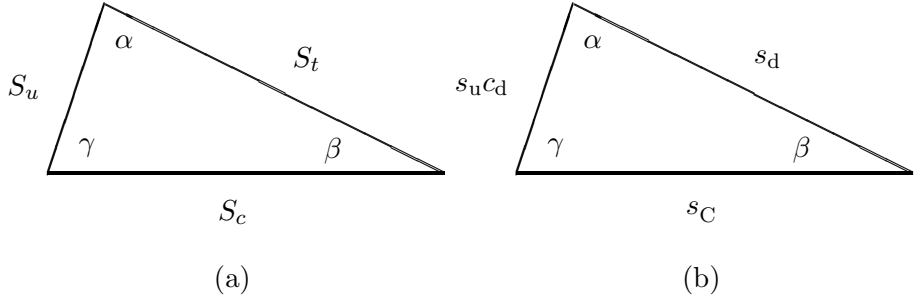
\begin{figure}[t]
\hspace*{-1.8cm}\begin{picture}(400,160)(10,210)
%-------------------------- (a)
\put(80,300){\line(1,0){150}}
\put(80,300.5){\line(1,0){150}}
%\put(80,299.5){\line(1,0){150}}
\put(150,285.5){\makebox(0,0){$S_c$}}
\put(80,300){\line(1,3){21.5}}
\put(80,300.5){\line(1,3){21.5}}
\put(80,299.5){\line(1,3){21.5}}
\put(71,333){\makebox(0,0){$S_u$}}
\put(230,300){\line(-2,1){128}}
\put(230,300.5){\line(-2,1){128}}
%\put(230,299.5){\line(-2,1){128}}
\put(178,343.5){\makebox(0,0){$S_t$}}

\put(95,310){\makebox(0,0){$\gamma$}}
\put(188,309){\makebox(0,0){$\beta$}}
\put(109,350){\makebox(0,0){$\alpha$}}
\put(150,260){\makebox(0,0){(a)}}

%--------------------------(b)
\hspace*{-1.5cm}\put(300,300){\line(1,0){150}}
\put(300,300.5){\line(1,0){150}}
%\put(300,299.5){\line(1,0){150}}
\put(370,285.5){\makebox(0,0){$s^{~}_{\rm C}$}}
\put(300,300){\line(1,3){21.5}}
\put(300,300.5){\line(1,3){21.5}}
\put(300,299.5){\line(1,3){21.5}}
\put(287,333){\makebox(0,0){$s_{\rm u} c_{\rm d}$}}
\put(450,300){\line(-2,1){128}}
\put(450,300.5){\line(-2,1){128}}
%\put(450,299.5){\line(-2,1){128}}
\put(395,343.5){\makebox(0,0){$s_{\rm d}$}}

\put(315,310){\makebox(0,0){$\gamma$}}
\put(408,309){\makebox(0,0){$\beta$}}
\put(329,350){\makebox(0,0){$\alpha$}}
\put(370,260){\makebox(0,0){(b)}}
\end{picture}
\vspace{-1.cm}
\caption{The unitarity triangle (a) and its rescaled counterpart (b)
in the complex plane.}
\end{figure}
%%%%%%%%%%%%%%%%%%%%%%%%%%%%%%%%%%%%%%%%%%%

i) At future $B$-meson factories, the study of $CP$ violation will
concentrate on measurements of the unitarity triangle 
\begin{equation}
S_u ~ + ~ S_c ~ + ~ S_t \; = \; 0 \; ,
%               (23)
\end{equation}
where $S_i \equiv V_{id} V^*_{ib}$ in the complex
plane (see Fig. 2(a)). The inner angles of this triangle 
are denoted as 
\begin{eqnarray}
\alpha & \equiv & \arg (- S_t S^*_u ) \; , \nonumber \\
\beta  & \equiv & \arg (- S_c S^*_t ) \; , \nonumber \\
\gamma & \equiv & \arg (- S_u S^*_c ) \; .
%               (24)
\end{eqnarray}
In terms of the parameters
$\theta$, $\theta_{\rm u}$, $\theta_{\rm d}$ and $\varphi$, we obtain
\begin{eqnarray}
\sin (2\alpha) & = & \frac{2 c_{\rm u} c_{\rm d} \sin\varphi \left
( s_{\rm u} s_{\rm d} c + c_{\rm u} c_{\rm d} \cos\varphi \right )}{s^2_{\rm
u} s^2_{\rm d} c^2 + c^2_{\rm u} c^2_{\rm d} + 2 s_{\rm u} c_{\rm u} s_{\rm d} c_{\rm d} c
\cos\varphi} \; , \nonumber \\ \nonumber \\
\sin (2\beta) & = & \frac{2 s_{\rm u} c_{\rm d} \sin\varphi \left
( c_{\rm u} s_{\rm d} c - s_{\rm u} c_{\rm d} \cos\varphi \right )}{c^2_{\rm
u} s^2_{\rm d} c^2 + s^2_{\rm u} c^2_{\rm d} - 2 s_{\rm u} c_{\rm u} s_{\rm d} c_{\rm d} c
\cos\varphi} \; .
%               (25)
\end{eqnarray}
To an excellent degree of accuracy, one finds $\alpha \approx
\varphi$. In order to illustrate how accurate this relation is, let us
input the central values of $\theta$, $\theta_{\rm u}$ and $\theta_{\rm 
d}$ (i.e., $\theta = 2.25^{\circ}$, $\theta_{\rm u} = 4.57^{\circ}$
and $\theta_{\rm d} = 12.7^{\circ}$) to Eq. (25). Then one arrives at
$\varphi - \alpha \approx 1^{\circ}$ as well as $\sin (2\alpha)
\approx 0.34$ and $\sin (2\beta) \approx 0.65$. 
It is expected that $\sin (2\alpha)$ and $\sin (2\beta)$
will be directly measured from the $CP$ asymmetries in 
$B_d \rightarrow \pi^+\pi^-$ and $B_d \rightarrow J /\psi K_S$ modes
at a $B$-meson factory.

Note that the three sides of the unitarity triangle 
(23) can be rescaled by $|V_{cb}|$. In a very good approximation
(with the relative error $\sim 2\%$), one arrives at
\begin{equation}
|S_u| ~ : ~ |S_c| ~ : ~ |S_t| \; \approx \; s_{\rm u} c_{\rm d} ~ : ~ 
s^{~}_{\rm C} ~ : ~ s_{\rm d} \; .
%               (26)
\end{equation}
Equivalently, one can obtain
\begin{equation}
s_{\alpha} ~ : ~ s^{~}_{\beta} ~ : ~ s_{\gamma} \; \approx \; s^{~}_{\rm C} 
~ : ~ s_{\rm u} c_{\rm d} ~ : ~ s_{\rm d} \; ,
%               (27)
\end{equation}
where $s_{\alpha} \equiv \sin\alpha$, etc.
The rescaled unitarity triangle is shown in Fig. 2(b). Comparing this
triangle with the LQ triangle in Fig. 1, we find that they are 
indeed congruent with each other to a high degree of accuracy.
The congruent relation between these two triangles is particularly
interesting, since the LQ triangle is essentially a feature of the physics
of the first two quark families, while the unitarity triangle is
linked to all three families. In this connection it is of special
interest to note that in models which specify the textures of the mass 
matrices the Cabibbo triangle and hence three inner angles of the unitarity
triangle can be fixed by the spectrum of the light quark masses and
the $CP$-violating phase $\varphi$ (see, e.g., Ref. \cite{FX95}).

j) It is worth pointing out that the u-channel and d-channel mixing
angles are related to the so-called Wolfenstein parameters 
\cite{Wolfenstein83} in a simple way:
\begin{eqnarray}
\tan\theta_{\rm u} & = & \left | \frac{V_{ub}}{V_{cb}} \right | 
\; \approx \; \lambda \sqrt{\rho^2 + \eta^2} \; , \; \nonumber \\
\tan\theta_{\rm d} & = & \left | \frac{V_{td}}{V_{ts}} \right |
\; \approx \; \lambda \sqrt{ (1-\rho)^2 + \eta^2} \; ,
%               (28)
\end{eqnarray}
where $\lambda \approx s^{~}_{\rm C}$ measures the magnitude of $V_{us}$.
Note that the $CP$-violating parameter $\eta$ is linked to $\varphi$
through
\begin{equation}
\sin\varphi \; \approx \; \frac{\eta}{\sqrt{\rho^2 + \eta^2}
\sqrt{(1-\rho)^2 + \eta^2}} \; 
%               (29)
\end{equation}
in the lowest-order approximation. Then $\varphi =90^{\circ}$ implies
$\eta^2 \approx \rho ( 1- \rho)$, on the condition $0 < \rho < 1$. In
this interesting case, of course, the flavor mixing matrix can 
fully be described in terms of only three independent parameters.

k) Compared with the standard parametrization of the flavor mixing
matrix $V$ advocated in
Ref. \cite{GKR}, the parametrization (12) has an additional
advantage: the renormalization-group evolution of $V$, from the weak
scale to an arbitrary high energy scale, is 
to a very good approximation associated only with the angle $\theta$. This can
easily be seen if one keeps the $t$ and $b$ Yukawa couplings only  
and neglects possible threshold effect in the one-loop
renormalization-group equations of the Yukawa matrices.
Thus the parameters $\theta_{\rm u}$, $\theta_{\rm d}$ and $\varphi$
are essentially independent of the energy scale, while $\theta$ does
depend on it and will change if the underlying scale is shifted, say
from the weak scale ($\sim 10^2$ GeV) to the grand unified theory
scale (of order $ 10^{16}$ GeV). In short, the heavy quark mixing is
subject to renormalization-group effects; but the u- and d-channel
mixings are not, likewise the phase $\varphi$ describing $CP$
violation and the LQ triangle as a whole.

We have presented a new description of the flavor mixing 
phenomenon, which is based on the phenomenological fact that the quark 
mass spectrum exhibits a clear hierarchy pattern. This leads uniquely
to the interpretation of the flavor mixing in terms of a heavy quark
mixing, followed by the u-channel and d-channel mixings. The complex
phase $\varphi$, describing the relative orientation of the u-channel
mixing and the d-channel mixing in the complex plane, signifies
$CP$ violation, which is a phenomenon primarily linked to the physics
of the first two families. The Cabibbo angle is not a basic mixing
parameter, but given by a superposition of two terms involving the
complex phase $\varphi$. The experimental data suggest that the phase
$\varphi$, which is directly linked to the phases of the quark mass
terms, is close to $90^{\circ}$. This opens the possibility to
interpret $CP$ violation as a maximal effect, in a similar way as
parity violation. 

Our description of flavor mixing has many clear advantages compared
with other descriptions. We propose that it should be used in the
future description of flavor mixing and $CP$ violation, in particular, 
for the studies of quark mass matrices and $B$-meson physics.

The description of the flavor mixing phenomenon given above is of special
interest if for the u- and d-channel mixings the quark mass textures
discussed first in Ref. \cite{Fritzsch77} are applied (see also Ref. \cite{FX95}).
In that case one finds 
\begin{eqnarray}
\tan \theta_{\rm u} & = & \sqrt{\frac{m_u}{m_c}} \; , \nonumber
\\
\tan \theta_{\rm d} & = & \sqrt{\frac{m_d}{m_s}} \; , 
%               (30)
\end{eqnarray}
apart from small corrections \cite{FritzschXing98}.
The experimental value for $\tan \theta_{\rm u}$ given by the ratio
$|V_{ub} / V_{cb}|$ is in agreement with the observed value for
$\sqrt{m_u / m_c} \approx 0.07$, but the error bars for both
$m_u / m_c $ and $|V_{ub} / V_{cb}|$ are almost the same
(about $25\%$). Thus from the underlying texture no new information is
obtained.

This is not true for the angle $\theta_{\rm d}$, whose experimental value
is due to a large uncertainty: $\theta_{\rm d} = 12.7^{\circ } \pm 3.8^{\circ }$.
If $\theta_{\rm d}$ is given indeed by the square root of the quark mass ratio
$m_d /m_s$, which is known to a high degree of accuracy
\cite{Leutwyler96}, we would know $\theta_{\rm d}$ and 
therefore all four parameters of the flavor mixing matrix with high precision.

As emphasized in Ref. \cite{FX95}, the phase angle $\varphi $ is very close
to 90$^{\circ }$, implying that the LQ triangle and the
unitarity triangle are essentially rectangular triangles. In particular the
angle $\beta $, which is likely to be measured soon from the 
$CP$ asymmetry in  $B_d \rightarrow J / \psi K_S$, is expected to be close
to $20^{\circ }$.

It will be very interesting to see whether the angles $\theta_{\rm u}$ and
$\theta_{\rm d}$ are indeed given by the square roots of the light quark mass
ratios $m_u / m_c$ and $m_d /m_s$, which imply that the phase $\varphi$
is close to or exactly equal to $90^{\circ}$. This would mean that the light quarks
play the most important role in the dynamics of flavor mixing and $CP$
violation and that a small window has been opened allowing the first view
accross the physics landscape beyond the mountain chain of the Standard
Model.

\vspace{0.5cm}

{\bf Acknowledgements}: It is a pleasure to thank the organising team of the
University of Katowice, especially K.\ Kolodziej, for the efforts to
organise this Physics School at Ustro\a'{n} in the Beskide mountains.

\end{document}